\documentclass[runningheads,a4paper]{llncs}

\usepackage{amssymb}
\usepackage{longtable}
\usepackage{array}
\usepackage{graphicx}
\usepackage{makecell,multirow}
\usepackage[numbers]{natbib}
\usepackage{url}
\usepackage{enumitem}

\usepackage{etoolbox} 
\AtBeginEnvironment{longtable}{}

\urldef{\mailsa}\path|{bsaha, mhasan12, nanjum, stahora, asiddika}@students.kennesaw.edu|
\urldef{\mailsc}\path|hshahria@kennesaw.edu|    
\newcommand{\keywords}[1]{\par\addvspace\baselineskip
\noindent\keywordname\enspace\ignorespaces#1}

\def\BibTeX{{\rm B\kern-.05em{\sc i\kern-.025em b}\kern-.08em
    T\kern-.1667em\lower.7ex\hbox{E}\kern-.125emX}}
\newcolumntype{P}[1]{>{\compress\vspace{-2ex}}p{#1}<{\vspace*{-2ex}}}

\begin{document}

\mainmatter  

\title{Protecting the Decentralized Future:\\An Exploration of Common Blockchain Attacks and their Countermeasures}

\titlerunning{Protecting the Decentralized Future}
\authorrunning{Bilash S et al.}

%
%
\author{Bilash Saha*%
\and Md Mehedi Hasan \and Nafisa Anjum \and Sharaban Tahora  \and Aiasha Siddika
\and Hossain Shahriar}
%

\institute{{Department of Information Technology, \\
Kennesaw State University, Georgia, United States \\}
\mailsa\\
\mailsc\\}

%
%
\maketitle

\begin{abstract}
Blockchain technology transformed the digital sphere by providing a transparent, secure, and decentralized platform for data security across a range of industries, including cryptocurrencies and supply chain management. Blockchain's integrity and dependability have been jeopardized by the rising number of security threats, which have attracted cybercriminals as a target. By summarizing suggested fixes, this research aims to offer a thorough analysis of mitigating blockchain attacks. The objectives of the paper include identifying weak blockchain attacks, evaluating various solutions, and determining how effective and effective they are at preventing these attacks. The study also highlights how crucial it is to take into account the particular needs of every blockchain application. This study provides beneficial perspectives and insights for blockchain researchers and practitioners, making it essential reading for those interested in current and future trends in blockchain security research.
\end{abstract}

\section{Introduction}

Blockchain technology has the potential to revolutionize various industries, from finance to healthcare to supply chain management. The decentralized nature of blockchain networks offers several advantages over traditional centralized systems, such as increased transparency, immutability, and security. According to ‘Deloitte’s 2021 global blockchain survey’ \cite{deloitte2021blockchain}, 76\% respondents stated that their companies adopted blockchain as a digital asset, and 83\% of them believe that digital assets will alternate fiat currency in the next 10 years. A report by Grand View Research Inc. stated that by 2030 global blockchain technology market size is expected to extend to 1,431.54 billion USD with a compound annual growth rate of 87.7\% from the year 2023 to 2030 \cite{prnewswire2022global}. In line with this trend, a report titled, “Blockchain Distributed Ledger Market by Component, Type, Enterprise Size, Application and End User, Opportunity Analysis and Industry Forecast, 2020–2027”, the global blockchain ledger market size is valued at USD 2.89 billion and by 2027 extrapolated to outstretch USD 137.29 billion at a CAGR of 62.7\% growing rate \cite{blockchainmarket}. The financial services sector is leading the pack, accounting for a substantial share of the blockchain market, as per the report. Following the success of bitcoin, a multitude of diverse applications of blockchain has grown rapidly including cross-border money transfer \cite{9036283}, smart contract supported by EVM (Ethereum virtual machine) \cite{8643084}, adaptation into IoT ecosystem \cite{s18082575, saha2023blockthefall}, data and identity security \cite{tahora2023blockchain, shrier2016blockchain}, electronic healthcare records (EHRs) \cite{holbl2018systematic, randolph2022blockchain, faruk2022security}, automated logistics \cite{8493157} and non-fungible tokens (NFTs) \cite{9877874}.

However, these advantages also come with unique challenges. Blockchain networks are susceptible to various attacks that can compromise their security and integrity. According to NIST, 20139 vulnerabilities were published in 2021 \cite{byers_national_2022}, and SonicWall reported an extreme rise in cybercrimes in the 2022 cyber threat report \cite{SonicWall22}. Till now there have been numerous reported attacks on blockchain systems, including the most common 51\% attacks, double-spending attacks, and smart contract vulnerabilities. For example, Ethereum Classic has suffered multiple 51\% double-spending attacks in 2019 and 2020, three repeated attacks in a month, resulting in the theft of more than \$6.7 million \cite{badertscher2021rational, sinclair_ethereum_2020}. On August 10, 2021, an attacker hacked a smart contract of the Poly Network and stole \$611 million in cryptocurrency, the largest crypto-related hack to date \cite{9936330}. The largest cryptocurrency exchange, Binance, suffered a major attack in October 2022, with hackers stealing 2 million BNB tokens through a blockchain vulnerability and causing a \$570 million loss \cite{livni_binance_2022, oxford2022binance}. These recent attacks demonstrate that blockchain systems are highly vulnerable to cyber-attacks.

As the usage of blockchain technology grows, it becomes crucial to understand these attacks and how to prevent them. A critical analysis of malicious activities within blockchain technologies and their consequential influences can facilitate practitioners to respond to security concerns from an early development phase. Thus, to shape the research and formulate a strategic agenda, we offered the following  research questions:
\begin{itemize}
\item RQ1: What are the most common types of attacks on blockchain technology and how do they impact the security and integrity of the system?

\item RQ2: What security measures (i.e., what methods or technologies) have been employed to detect and mitigate malicious blockchain attacks?
\end{itemize}

To answer the above-mentioned questions, we have conducted a comprehensive qualitative and quantitative approach on existing studies regarding blockchain privacy and security issues so that this research can be presented as a potential resource for offering a holistic outline and defending current and forthcoming attacks.

\section{Method}
We emphasize the defining criteria we used, our structured literature search procedure, and the process we underwent for analyzing the data.

\subsection{ Scoping Criteria}
The scoping criteria for the research questions about the most common types of attacks on blockchain technology, their impact on security and integrity, and the security measures employed to detect and mitigate malicious attacks encompass a comprehensive analysis of the various aspects of blockchain security. The research delves into the various attack methods used to compromise the security of blockchain systems, including the 51\% attack, double-spend attack, smart contract vulnerability attack, and Sybil attack. The 51\% attack, for instance, occurs when an attacker gains control over more than half of the computational power of the network, allowing them to manipulate transactions and potentially reverse previous transactions, as seen in the Ethereum Classic blockchain in 2019. The study also evaluates network security protocols like firewalls and intrusion detection systems, security audits conducted by outside security experts, and consensus algorithms like Proof-of-Work (PoW) and Proof-of-Stake (PoS) to detect and prevent malicious attacks on blockchain. Cryptography techniques like digital signatures are also evaluated. The study examines how these safeguards have changed over time and how they have been modified to counter evolving blockchain attacks. We also look at how blockchain attacks have developed over time, including how attack techniques and motivations have changed. For instance, early attacks on blockchain systems were primarily concerned with taking advantage of technical flaws, whereas more recent attacks have evolved and become more sophisticated with monetary gain as their primary motivation. The study looks at the security measures created to counteract these evolving attack techniques and goals. The research considers what can be anticipated in the future with regard to the evolution of blockchain attacks and security measures. Finally, it addresses the current challenges in the field of blockchain security. It is crucial to continuously assess and update the security measures used to protect blockchain systems because as blockchain technology develops, new attack vectors and defenses are likely to appear.

\begin{figure*}[ht]
\centerline{\includegraphics[width=1.0\textwidth]{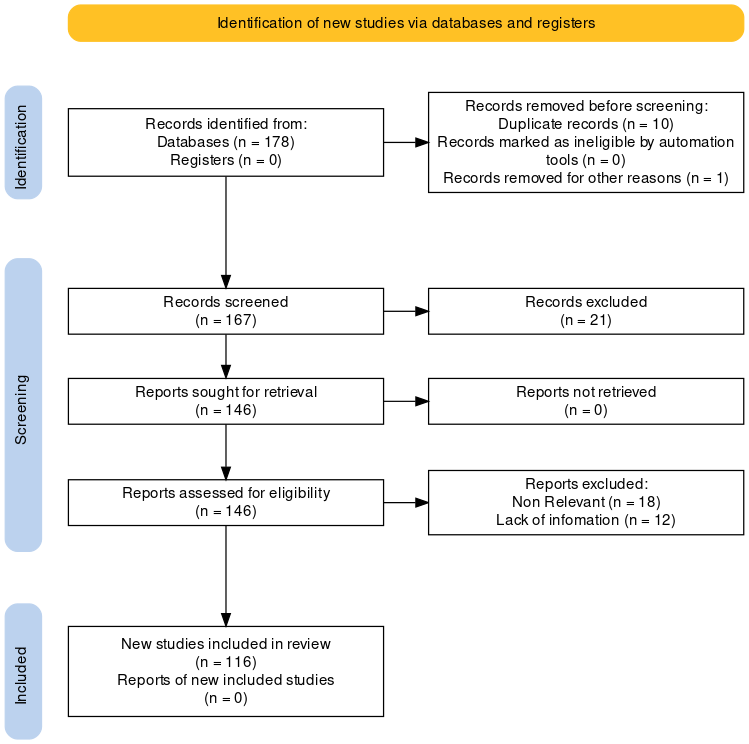}}
\caption{ PRISMA flow diagram for study selection in the systematic review }
\label{fig:prisma}
\end{figure*}

\subsection{Systematic Literature Search}
A structured and methodical approach to finding, reviewing, and synthesizing pertinent research reports, articles, and other academic publications on a particular subject is known as a systematic literature search. The terms "blockchain security," "attack types," "51 percent attack," "double-spend attack," "smart contract vulnerability," "Sybil attack," "consensus algorithms," "cryptography," "network security protocols," "security audits," "blockchain evolution," "current challenges," and "future trends" are used in research questions about the most typical types of attacks on blockchain technology, their effects on security and integrity, and the security measures used to protect against them. The search aims to identify relevant articles, reports, and studies in academic databases, for instance, Google Scholar, IEEE Xplore, and ACM Digital Library, as well as industry sources, such as CERTs, government publications, and online forums. The search considers studies published in the last decade and applies filters, such as publication type, language, and date, to narrow down the search results. The resulting studies are then critically evaluated and synthesized to address the research questions and provide a comprehensive analysis of the state of blockchain security. After thoroughly reviewing 178 research papers, we carefully selected 116 papers that were most relevant to our research. The PRISMA flowchart utilized for study selection is shown in Figure \ref{fig:prisma}.

\section{Results}

\subsection{Attacks on Blockchain Systems and the Reasons Behind Their Occurrences}
Blockchain technology is generally utilized in numerous sorts of applications like healthcare, farming, smart grid, smart city, commercial approach, etc. due to its privacy and security attributes. Nevertheless, there are some highly sophisticated security attacks that have the potential to cause huge irreversible destruction. In this section, we will introduce the various and most common types of security attacks leveraged by cyber attackers destabilizing the practice of blockchain networks, i.e., 51\% attack, double-spending attack, replay attack, sybil attack, eclipse attack, and so forth.

\subsubsection{51\% Attack}
Many of the blockchain systems suffer from a 51\% attack, also known as a majority attack, that is considered a high risk for the integrity and security of these blockchains, where an attacker or a group of miners control more than 50\% of the network's computing power, allowing them to manipulate the consensus mechanism and double-spend coins. One of the main indications of a 51\% attack is when the network behavior suddenly changes. By executing a 51\% attack, an attacker can autocratically modify the information on the blockchain system. With the assistance of such exploitation, an attacker can perform the following attacks: a) reverse transactions and initiation of a double-spending attack which means spending the exact coins multiple times; b) modify the order of transactions; c) hamper the usual activities of other miners; and d) prohibit the confirmation activity of regular transactions.
\subsubsection{Smart Contract Vulnerability}
Smart contracts are self-executing contracts automatically executed when specific conditions are met. Such vulnerabilities can lead to security breaches, such as the loss of funds or unauthorized access to confidential information. It is a decentralized program on a blockchain network that automates and enforces agreements without a trusted third party and contains preset conditions for auto-execution and is transparent, immutable, and publicly visible \cite{luu2016making, peng2021security, sayeed2020smart}. Smart contract vulnerabilities refer to security issues in smart contracts, often arising from coding mistakes or design flaws \cite{mense2018security}. These vulnerabilities can lead to unintended consequences and are difficult to fix due to the inherent immutability of the blockchain \cite{mense2018security}.

\begin{figure*}[ht]
\centerline{\includegraphics[width=1.0\textwidth]{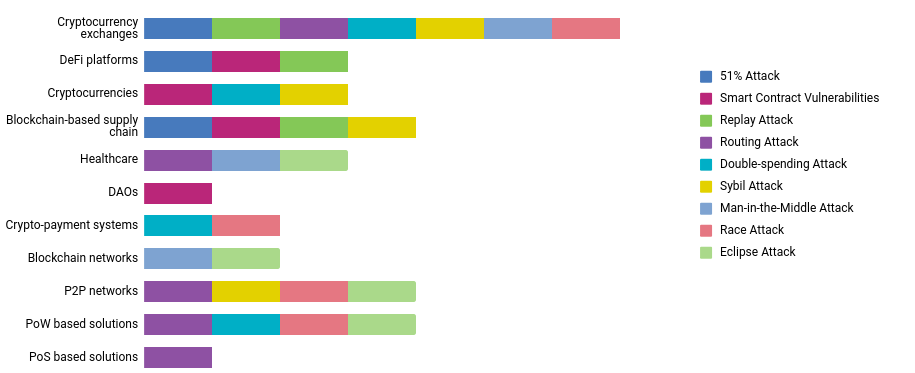}}
\caption{ Blockchain Attacks and impacted sectors }
\label{fig:impacted_sectors}
\end{figure*}

\subsubsection{Double-Spending Attack}
Double-spending occurs when the same digital currency or asset is spent twice, leading to a data consistency attack. In one recent attack of January 2021, the attacker was able to double spend \$3.3 million worth of cryptocurrency on the blockchain of the privacy-focused cryptocurrency, Monero (XMR). A prevalent weakness that may enable a double spending attack is when a cryptocurrency network is dependent on a consensus algorithm that lacks adequate security or rapidity in verifying transactions within a timely window and fails to store older blocks before removing them \cite{begum2020blockchain, marcus2018low}. This creates a potential opportunity for an adversary to deceitfully repeat the spending of a transaction, exploiting the lag in the network's detection mechanism. Another study states that the confirmation time required for a transaction to undergo validation and inclusion into the blockchain creates the potential vulnerability for the attacker to attempt to expend the same units of cryptocurrency once more, thereby triggering a double-spending scheme \cite{deirmentzoglou2019survey}. Additionally, the acceptance of insecure and unconfirmed transactions, coupled with the direct connection of incoming nodes to the main chain, creates an additional vulnerability in the blockchain system that can facilitate double spending \cite{grundmann2019exploiting}.
\subsubsection{Man-in-the-middle Attack}
In a blockchain system, a Man-in-the-Middle (MITM) attack is a type of attack that can occur when an attacker intercepts and modifies the communication between two nodes in a blockchain network \cite{riadi2022block}. The attacker may be able to alter the data being transmitted, steal private keys, or tamper with transaction data to benefit themselves \cite{razmjouei2020ultra}. This can cause significant damage to the blockchain network, leading to loss of funds, data breaches, and other security risks \cite{akter2020man}.
\subsubsection{Routing Attack}
In a blockchain network, the nodes communicate with each other to verify transactions, maintain the network's security, and reach a consensus on the state of the ledger. The nodes use a routing protocol to communicate with each other and determine the best path for propagating transactions and blocks. A Routing Attack occurs when an attacker tries to manipulate the routing protocol to control the flow of information in the network, leading to the redirection of traffic and the compromise of nodes or data \cite{aggarwal2021attacks}. The attacker may do this by sending false routing information to the nodes, which can lead to the nodes selecting a malicious path for propagating transactions and blocks \cite{aggarwal2021attacks}. This can allow the attacker to carry out various attacks, such as double-spending attacks, denial of service attacks, and 51\% attacks \cite{careem2020reputation}.
\subsubsection{Sybil Attack}
A Sybil attack is an attack on a peer-to-peer network where a malicious node creates multiple fake identities to disrupt the network. It can be effective, particularly in consensus mechanisms utilizing Proof-of-Work (PoW), where a node's computational power governs its capacity to authenticate transactions and earn rewards. In May 2022 the Ethereum-based DeFi protocol, Saddle Finance, suffered a Sybil attack that exploited vulnerabilities in its on-chain governance system and lost more than \$10 million \cite{noauthor_saddle_2022}. A comparable Sybil attack occurred in August 2021, resulting in the Liquid Global cryptocurrency exchange being breached for a staggering sum of \$90 million \cite{coindesk-liquid-global-hack}. Another study highlighted that the node authentication process of the P2P network during the blockchain network's joining phase poses a significant vulnerability to the system, making it susceptible to the insertion of Sybil nodes that may result in block propagation delays.

\subsubsection{Race Attack}
One of the most preeminent means of attack which enables various double-spending attacks is denoted as a race attack wherein a threat actor initiates two contradictory transactions into the network concurrently \cite{shemov2020blockchain, averin2019review}. The prime transaction is dispatched to the victim who agrees to the transaction, accepts the payment, and dispatches the product without awaiting confirmation. Instantaneously, the remaining contrasting transaction is broadcasted to the network retracting the same amount of cryptocurrency and eventually making the first transaction corrupted \cite{konig2020risks}. If both recipients neglect to authenticate first, this enables the attacker to obtain twofold quantities of products for an equal amount of money \cite{rathod2018security}.

\subsubsection{Eclipse Attack}
An eclipse attack occurs when a malicious actor controls all the peers that a targeted node is connected to, allowing them to steal information, compromise the targeted node and establish a degree of isolation of the earmarked nodes from the entire P2P network by blocking communication allowing only the attacking nodes to receive the details \cite{castro2002secure, aggarwal2021attacks}. Typically, this sort of attack entails sending a transaction to a targeted node, hence showing proof of a payment, and eclipsing it from the network. The final step involves delivering another transaction to the whole network that effectively spends the same tokens again \cite{castro2002secure}. This sequence of actions takes over the network by monopolizing and exerting control over its connections.

\subsubsection{Replay Attack}
A replay attack is a type of network attack where a malicious actor intercepts, captures, and resends valid digital transaction data, thus exploiting the need for the original data to be validated and stealing funds or disrupting the network \cite{nfting_what_2022}. This can occur when blockchain ledgers undergo hard forks or chain splits, resulting in separate ledgers that share transaction history. The cause of vulnerability in this scenario is due to incompatible transaction or message formats, shared transaction history, and the lack of replay protection \cite{duan2022multiple}. Replay attacks in the blockchain space can significantly impact various sectors, including cryptocurrency exchanges, decentralized finance (DeFi), and blockchain-based supply chain systems \cite{nfting_what_2022}.

Table \ref{tab:tab1} represents these security attacks on blockchain mentioned earlier and provides a detailed overview of different tactics and techniques employed by cyber attackers that can cause such vulnerabilities resulting in devastating consequences in countless sectors.


\begingroup
\setlist[itemize]{label={--},nosep, leftmargin={0.6cm}, before=\vspace*{-\baselineskip}, after=\vspace*{-0.9\baselineskip}}
\setlength{\extrarowheight}{3pt}
\begin{small}
\begin{longtable}[ht!]{|>{\centering\arraybackslash}m{1.5cm}|>{\centering\arraybackslash}m{5.7cm}|>{\centering\arraybackslash}m{3.7cm}|>{\centering\arraybackslash}m{1.cm}|}

\caption{A comprehensive overview of attacks on blockchain technology} 
\label{tab:tab1}
\\

\hline 
\textbf{Attack Name} & \textbf{Cause of Vulnerability} & \textbf{Impact on Sectors} & \textbf{Refer- ences} \\
\hline

\endfirsthead

\multicolumn{4}{c}%
{{\bfseries \tablename\ \thetable{} -- continued from previous page}} \\
\hline
\textbf{Attack Name} & \textbf{Cause of Vulnerability} & \textbf{Impact on Sectors} & \textbf{Refer- ences} \\
\hline
\endhead
\hline \multicolumn{4}{|r|}{{Continued on next page}} \\ \hline
\endfoot
\hline
\endlastfoot

 51\% Attack & 
 \begin{itemize} 
\item Low network hash rate
\item Availability of hardware
\item Centralization of mining power
\item Renting hash power
\item Lack of economic incentives to behave honestly
\item Low cost of attack
\end{itemize} & 
\begin{itemize} 
    \item Cryptocurrency exchanges
    \item DeFi platforms
    \item Blockchain-based supply chain
\end{itemize} & 
\cite{aponte202151, shanaev2019cryptocurrency, saad2019overview} \\ \hline

Smart Contract Vulnerabilities & 
\begin{itemize} 
    \item Programming errors
    \item Incomplete understanding of the platform
    \item Complex code and business logic
    \item Inadequate testing and auditing
    \item Poor design choices
    \item Language and compiler limitations
    \item Reuse of vulnerable code
    \item Gas and resource management
    \item Lack of standards and best practices 
\end{itemize} & 
\begin{itemize} 
    \item DeFi platforms
    \item Cryptocurrencies
    \item Blockchain-based supply chain
    \item DAOs
\end{itemize} & 
\cite{praitheeshan2019security, vivar2020analysis} \\ \hline

Man-in-the-middle Attack & 
\begin{itemize} 
\item Intercepting the communication between two nodes in a blockchain network
\item Altering the transmitted data, stealing private keys, or tampering with transaction data. 
\end{itemize} & 
\begin{itemize} 
    \item Blockchain Networks
    \item Cryptocurrency exchanges
    \item Healthcare
\end{itemize} & \cite{riadi2022block, razmjouei2020ultra, akter2020man} \\
\hline

Replay Attack & \begin{itemize} \item Incompatible transaction/messages formats
\item Shared transaction history
\item  Lack of replay protection
\end{itemize} & 
\begin{itemize} 
    \item Cryptocurrency exchanges
    \item DeFi platforms
    \item Blockchain-based supply chain
\end{itemize} & \cite{sonnino2020replay, duan2022multiple} \\
\hline

Double-spending Attack & 
\begin{itemize} 
    \item Removing the older block while adding a new large chain
    \item Delay the propagation of correct block information
    \item Fake blocks initiated faster than valid ones using a fork
    \item Attackers created the blocks in advance
    \item Confirmation time
    \item Insecure and unconfirmed transaction acceptance
    \item Direct incoming node connection established with the main chain 
\end{itemize} & 
\begin{itemize} 
    \item Cryptocurrency exchanges
    \item PoW-based solutions
    \item Cryptocurrencies
    \item Crypto-payment systems
\end{itemize} & 
\cite{begum2020blockchain, zhang2019double, deirmentzoglou2019survey, marcus2018low, grundmann2019exploiting} \\
\hline

Routing Attack & \begin{itemize} \item Manipulation of the routing protocol to control the flow of information in the network
\item Sending out false routing information to the nodes 
\end{itemize} & \begin{itemize} 
    \item P2P networks
    \item PoW-based solutions
    \item PoS-based solutions
    \item Cryptocurrency exchanges
    \item Healthcare
\end{itemize} & \cite{aggarwal2021attacks, apostolaki2017hijacking} \\
\hline

Sybil Attack & \begin{itemize} \item Node reputation tampered by calculating reputation score combining with Sybil nodes
\item Sybil node insertion in the routing table (RTI attack)
\item Assuming the participating nodes are verified by consensus protocol
\item Building a profile by pairing with the same user (s) multiple times
\item Ignoring node authentication and less computing power
\end{itemize} & 
\begin{itemize} 
    \item P2P networks
    \item Cryptocurrency exchanges
    \item Cryptocurrencies
    \item Blockchain-based supply chain
\end{itemize} & 
\cite{otte2020trustchain, pradhan2018blockchain, rajab2020feasibility, quintyne2019short, swathi2019preventing, grundmann2019exploiting} \\
\hline

Race Attack & 
\begin{itemize} \item Unverified or duplicate transaction resulting in transactional fraud
\end{itemize} & 
\begin{itemize} 
\item P2P networks
\item Crypto-payment systems
\item PoW-based solutions
\item Cryptocurrency exchanges
\end{itemize} & 
\cite{morganti2018risk, conti2018survey, shemov2020blockchain} \\
\hline

Eclipse Attack & 
\begin{itemize} \item Insecure Internet Core Protocols and network-level liabilities allowing victims to be connected only to attacker-controlled peers
\item Ethereum’s acceptance of the Kademlia peer-to-peer practices can create numerous Ethereum nodes with ECDSA generating algorithms and only two machines with one IP address each
\item  Adversaries controlling the majority of the computational power in a PoW blockchain monopolizing connections transactions.
\item  Untrusted peers within public communities maneuvering honest nodes approach to the conventional blockchain network
\end{itemize} & 
\begin{itemize} 
\item PoW-based solutions
\item Healthcare
\item P2P networks
\item Blockchain Networks
\end{itemize} & 
\cite{alangot2021decentralized, marcus2018low, xu2020eclipsed, dai2022eclipse, signorini2018advise, heilman2015eclipse} \\
\hline
\end{longtable}
\end{small}
\endgroup

\subsection{Detect and Mitigate Malicious Blockchain Attacks}
As the adoption of blockchain technology continues to grow, so too does the potential for malicious attacks. Consequently, various methods and technologies have been developed to detect and mitigate these threats, enhancing the security and resilience of blockchain networks. From utilizing machine learning to identify suspicious activity to implementing robust consensus algorithms to prevent double-spending, these measures serve as the defensive wall against malicious attacks. Each subsection will focus on a specific type of malicious blockchain attack, offering a comprehensive understanding of their detection and mitigation.

\subsubsection{51\% Attack}
\subsubsection{Smart Contract Vulnerability}
Various vulnerability detection techniques have been employed to identify vulnerabilities in smart contracts. Static analysis methods include symbolic execution, where code is executed using symbols instead of actual values, allowing for the generation of algebraic terms and propositional formulas to analyze the code logic \cite{samreen2021survey, 9236931, e22020203, 10034747}. Control flow graph (CFG) construction represents the flow of execution as a directed graph, enabling a better understanding of the program's structure \cite{samreen2021survey, 9236931, e22020203}. Pattern recognition and rule-based analysis involve com code against known secure or vulnerable patterns and predefined rules to identify potential issues \cite{9236931, e22020203}. Decompilation analysis involves translating lower-level code into a higher-level representation to facilitate parsing and data flow analysis \cite{samreen2021survey}. Formal verification methods \cite{10034747} use mathematical proofs to verify that a program meets its specific properties.

Dynamic analysis techniques include execution trace at runtime, which captures the sequence of instructions executed during a particular run of the code \cite{samreen2021survey, e22020203}. Fuzzing input generation is an automated method that tests program execution by providing structured data as inputs and monitoring for unexpected behavior, such as unusual code paths or crashes \cite{samreen2021survey, 9236931, 10034747}.

To mitigate smart contract vulnerabilities and enhance security, several security measures can be taken based on established techniques and research papers \cite{samreen2021survey, 9236931, e22020203, 10034747}. Firstly, utilizing established design patterns and best practices can reduce the likelihood of introducing flaws during development. Implementing robust access control mechanisms ensures that only authorized parties can interact with the contract, mitigating risks associated with unauthorized access. Conducting regular code reviews and security audits can help identify potential vulnerabilities. Comprehensive testing, including static and dynamic analysis, fuzz testing, and formal verification methods \cite{samreen2021survey, e22020203, 10034747}, is essential to uncover vulnerabilities before deployment. Utilizing upgradeable contracts can enable developers to modify and improve contract functionality in response to discovered issues. Finally, continuous monitoring of deployed contracts and responding swiftly to detected vulnerabilities ensures that potential threats can be addressed proactively, further safeguarding smart contracts and the assets they manage.
\subsubsection{Man in the Middle Attack}
Network monitoring is an essential technique that can help detect unusual network activity, such as unexpected data transmissions or unexpected nodes joining the network \cite{choi2021blockchain}. It can also help identify any rogue nodes that are attempting to participate in the network \cite{progga2021securing}. Consensus checks can detect discrepancies in the transaction data being transmitted between nodes \cite{choi2021blockchain}. If a node detects a discrepancy, it can halt the transaction process and notify the network of the potential MITM attack \cite{wazid2020private, kulkarni2019preventing}. Digital signatures can help detect changes to the transaction data by ensuring that the data has not been tampered with during transmission \cite{kulkarni2019preventing, momeni2019fault}. Digital signatures can also help authenticate the transaction's sender and prevent unauthorized access  \cite{majeed2022coverage}. Reputation systems can be used to identify potentially malicious nodes in the network based on their previous behavior \cite{momeni2019fault, abdulrazzaq2020decentralized, dixit2022survey}. Nodes with a low reputation score can be flagged and prevented from participating in the network.

Encryption is an essential tool that can help prevent attackers from intercepting and understanding the data being transmitted between nodes \cite{homoliak2019security, rathod2022blockchain}. Encryption can be implemented at both the network and transaction levels, making it more challenging for attackers to access and steal sensitive data \cite{banu2021detection}. According to Kebande et al. and Rahman 2021 et al. Security \cite{kebande2021blockchain, rahman2021security}, multi-factor authentication can be used to protect access to private keys and stop unauthorized access by attackers. Multi-factor authentication can include techniques like biometric authentication, smart cards, and one-time passwords\cite{homoliak2019security, obaidat2020hybrid, ali2021secure}. Consensus mechanisms that require multiple nodes to verify and approve transactions before they are added to the blockchain can be created to thwart MITM attacks \cite{alkaeed2020distributed, yakubu2023blockchain, homoliak2019security}. As a result, manipulating transaction data may become more difficult for attackers. Identity management can be used to make sure that nodes on the network are authorized and authenticated before they are permitted to participate in the network\cite{ruf2021security}. Digital certificates, public-key infrastructure, and biometric authentication are a few examples of what can be done in this regard \cite{jurcut2020security, bello2020technical}. An extensive network monitoring program can assist in real-time detection and mitigation of potential MITM attacks\cite{abdallah2023extensive}. In order to identify and address any potential security risks before they seriously harm the network, this can include both network monitoring and transaction monitoring.


\begingroup
\setlist[itemize]{label={--},nosep, leftmargin={0.6cm}, before=\vspace*{-\baselineskip}, after=\vspace*{-0.9\baselineskip}}
\setlength{\extrarowheight}{3pt}
\begin{small}
\begin{longtable}[ht!]{|>{\centering\arraybackslash}m{1.5cm}|>{\centering\arraybackslash}m{4.0cm}|>{\centering\arraybackslash}m{5.5cm}|>{\centering\arraybackslash}m{1.0cm}|}
    
    \caption{A summarized overview of detection and mitigation techniques of blockchain attacks} 
    \label{tab:tab2}
    \\ \hline

    \textbf{Attack} & \textbf{Detection Technique} & \textbf{Security Measure (Mitigation)} & \textbf{Refer- ences} 
    \\ \hline
    \endfirsthead
    
    \multicolumn{4}{c}%
    {{\bfseries \tablename\ \thetable{} -- continued from previous page}} \\
    \hline
    \textbf{Attack} & \textbf{Detection Technique} & \textbf{Security Measures(Mitigation)} & \textbf{Refer- ences} \\
    \hline
    \endhead
    \hline \multicolumn{4}{|r|}{{Continued on next page}} \\ \hline
    \endfoot
    \hline
    \endlastfoot
    
    51\% Attack & 
    \begin{itemize} 
        \item Unusual Network Behavior
        \item Mined Blocks
        \item Block Reorganization
        \item Double-spending
        \item Suspicious Transactions
        \item Mining Pools
        \item Alerting Mechanisms
    \end{itemize} &
    \begin{itemize} 
        \item Implementing a more robust consensus mechanism
        \item Increasing the size of the network
        \item Using checkpoints
        \item Creating a decentralized network structure
        \item Creating a diverse community
    \end{itemize} &
    \cite{frankenfield201951, ye2018analysis, bastiaan2015preventing, shrestha2019regional, bae2018random} 
    \\ \hline

    Smart Contract Vulnerabilities & 
    \begin{itemize} 
        \item Smart Contract vulnerabilities
        \item Statically analyze the source code or bytecode of smart contracts
        \item Dynamic analysis: execute smart contracts with various inputs and monitor their runtime behavior
        \item Symbolic execution
        \item Formal verification \& Fuzz testing
        \item Code review and auditing
    \end{itemize} &
    \begin{itemize} 
        \item Use established design patterns and best practices
        \item Implement access control mechanisms
        \item Conduct regular code reviews and security audits
        \item Perform comprehensive testing
        \item Utilize upgradeable contracts
        \item Monitor and respond to detected vulnerabilities
    \end{itemize} &
    \cite{samreen2021survey, 9236931, e22020203, 10034747, praitheeshan2019security, vivar2020analysis}
    \\ \hline

    Double-spending Attack & 
    \begin{itemize} 
        \item Verification time
        \item Blocking time of transaction
        \item Constant monitoring of disconnection period
        \item Timestamp while building the chain
        \item Transaction verification
        \item Block confirmation
        \item Ledger confirmation
        \item Monitor node connection
        \item Updating the block rather than removing older one
    \end{itemize} &
    \begin{itemize} 
        \item Recipient-oriented transaction
        \item increase confirmation numbers
        \item Creating a consistent chain
        \item Longest chain rule for PoS protocols
        \item Key evolving cryptography
        \item Number of confirmed blocks expansion
        \item Disable unconfirmed connections
        \item Transaction forward
    \end{itemize} &
    \cite{begum2020blockchain, lee2018recipient, DBLP:journals/corr/abs-1805-05004, 8733108, 8653269, cryptoeprint:2018/236, 10.1007/978-3-662-58820-8_9}
    \\ \hline

    Race Attack & 
    \begin{itemize}
        \item Listening Period
        \item Insertion of Observers
    \end{itemize} &
    \begin{itemize} 
        \item Periodic transaction monitoring
        \item Vendors can insert observers nodes to transmit all transactions to himself
    \end{itemize} &
    \cite{shemov2020blockchain, dasgupta2019survey}
    \\ \hline

        Replay Attack & 
    \begin{itemize}
        \item Monitoring for duplicate transactions
        \item Analyzing transaction history in both chains
    \end{itemize} &
    \begin{itemize} 
        \item Implement Strong replay protection
        \item Splitting coins
        \item Opt-in Replay Protection
    \end{itemize} &
    \cite{sonnino2020replay, duan2022multiple, nfting_what_2022, dasgupta2019survey, islam2019blockchains}
    \\ \hline
    
    Sybil Attack & 
    \begin{itemize}
        \item Calculating node reputation and trustworthiness
        \item Node activity monitoring
        \item Monitoring node id generation in the initial phase of PoW
        \item Using social, computational, and monetary constraints
        \item Distributed behavior monitoring of each node
    \end{itemize} &
    \begin{itemize} 
        \item Permissionless, the tamper-proof data structure to store transactions
        \item Creating parallel transaction record block locally
        \item Address a registration server and calculate the reward score
        \item Restrict the unusually high computing power of nodes and increase the number of shards
        \item Burning currency to create a non-refundable deposit
        \item Time-locking to restrict the fund usage
        \item Fidelity bonds and coin-age
    \end{itemize} &
    \cite{10.1016/j.future.2017.08.048, 8710078, DBLP:journals/corr/abs-2002-06531, cryptoeprint:2019/1111, 8944507}
    \\ \hline

        Eclipse Attack & 
    \begin{itemize}
        \item Timestamp-based and gossip-based protocol
        \item Run seeding and eliminate the reboot exploitation window
        \item Anlomaly detection tools
    \end{itemize} &
    \begin{itemize} 
        \item Calculate average attack detection time and network load 
        \item Run the seeding procedure on the victim’s system
        \item Generate a threat database model from malicious forks and measures the abnormalities
    \end{itemize} &
    \cite{alangot2021decentralized, marcus2018low, xu2020eclipsed, signorini2018advise}
    \\ \hline

        Man-in-the-middle & 
    \begin{itemize} 
        \item detect unusual network activity
        \item detect discrepancies in the transaction data
        \item detects changes and authenticates access permission in the transaction data
        \item identify potentially malicious nodes in the network based on their behavior
    \end{itemize} &
    \begin{itemize} 
        \item Encryption of data being transmitted between nodes.
        \item Multi-factor Authentication to secure access to private keys
        \item verify and approve transactions before adding to the blockchain
        \item digital certificates, public-key infrastructure, and biometric authentication
        \item Continuous monitoring and response to potential attacks in real-time
    \end{itemize} &
    \cite{choi2021blockchain, kulkarni2019preventing, li2020lightweight, ahmad2020marine, homoliak2019security, yakubu2023blockchain}
    \\ \hline

    Routing Attack & 
    \begin{itemize}
        \item Network monitoring to track the routing paths of transactions and detect anomalies in the network traffic
        \item Data analysis to identify patterns and trends in the network
        \item Reputation systems to identify malicious nodes based on their behavior
    \end{itemize} &
    \begin{itemize} 
        \item Secure routing protocols to secure the routing paths in the network using digital certificates
        \item Utilizing distributed consensus mechanisms
        \item Employing peer-to-peer networking by allowing nodes to communicate directly with each other
        \item Node reputation systems towards identifying compromised nodes and prevent them from propagating transactions in the network
    \end{itemize} &
    \cite{sahay2020novel, tran2020stealthier, tekiner2021sok, arisdakessian2022survey}
    \\ \hline


\end{longtable}
\end{small}
\endgroup

\subsubsection{Routing Attack}
As it can be difficult to distinguish between safe network behavior and malicious behavior, detecting a routing attack can be difficult. To identify routing attacks in a blockchain, there are a number of techniques that can be used. Tracking transaction routing paths and spotting anomalies in network traffic are both possible with network monitoring \cite{sahay2020novel}. A Routing Attack, for instance, may be indicated by a transaction propagating through an unusual path \cite{tran2020stealthier}. Cite: tekiner2021sok, arisdakessian2022survey. Data analysis can be used to find patterns and trends in network traffic that might indicate a Routing Attack\cite{ tekiner2021sok, arisdakessian2022survey}. For instance, a Routing Attack may be indicated by a sudden increase in the number of transactions being propagated through a particular node \cite{abbassi2022bcsdn}. Reputation systems can be used to spot compromised or maliciously acting nodes. Each node can receive a reputation score from the reputation system based on how it behaves, and nodes with low reputation scores may be potential sources of routing attacks.

In the blockchain, there are a number of techniques that can be used to stop and lessen routing attacks. The routing paths in the blockchain network can be made secure by using secure routing protocols like BGPsec and RPKI \cite{mastilak2022secure}. These protocols use digital certificates to authenticate the routing information and prevent spoofing attacks. Distributed consensus mechanisms such as Proof of Work (PoW) and Proof of Stake (PoS) can be used to prevent Routing Attacks by making it difficult for attackers to control the majority of the network's computational power or stake \cite{sayeed2019assessing}. Peer-to-peer networking can be used to increase the resiliency of the network by allowing nodes to communicate directly with each other, rather than relying on a centralized routing authority \cite{ihle2023incentive}. This makes it harder for an attacker to manipulate the routing paths and control the flow of information in the network. Node reputation systems can be used to identify nodes that are behaving maliciously or are compromised, and prevent them from propagating transactions and blocks in the network \cite{de2020blockchain}. Nodes with a low reputation score can be prevented from participating in the network or be subject to additional scrutiny to prevent them from carrying out a Routing Attack.

\subsubsection{Race Attack}
Countless amount of comprehensive research has established a couple of detection measures that can be employed to identify this type of attack. A major detection technique is termed a Listening Period where each transaction received is associated with a listening timeframe by the vendor, with the aim of scrutinizing all transactional records during this period. If the vendor does not detect any effort of race attack within this listening period, he may proceed to deliver the product \cite{alsunbul2021blockchain}. Additionally, another detection method is the Insertion of Observers where a vendor can insert one or two nodes into the network referred to as “monitors” which function to transmit all transactions to the vendor, thereby facilitating the swift detection of any race attack or double-spending attack endeavors \cite{alsunbul2021blockchain}.

There exists several preventive measures that can be employed to mitigate the risk of race attacks, involving the implementation of time-stamping for all transactions to enable the recording of their processing time, thus allowing the prevention of race attacks by considering the first recorded transaction as valid. Additionally, cryptographic signatures can be applied to sign all the transactions in the blockchain network indicating that the valid transaction was submitted by the intended user. Multi-party signatures can also be leveraged to further reinforce transaction security by ensuring that multiple parties have signed off on the transaction, thus mitigating the risk of a race attack. Furthermore, increasing the processing time for each bock can also serve to prevent a race attack as it would require more time for an attacker to process multiple transactions within a restricted timeframe \cite{morganti2018risk}.

\subsubsection{Eclipse Attack}
A plethora of research has been conducted to establish detection approaches that can be employed in order to recognize eclipse attacks. The first strategy identifies eclipse attacks by utilizing the timestamps of questionable blocks where if the amount of time between newly created blocks is higher than usual it concludes that the network has been compromised. In this particular method, it reasonably requires approximately 2-3 hours to understand that they are being targeted by an eclipse attack \cite{alangot2021decentralized}. Another technique is termed “ubiquitous gossip protocol” in which the user rides along gossip messages on its associations to protocol-dominated servers without the necessity of managing any alterations to Bitcoin or the peer-to-peer network. The support from web servers essential to connect with Bitcoin block headers is also negligible \cite{anita2019blockchain}. Amongst other conceivable detection methods, an alternative study suggests imitating the keys on several nodes, thus enumerating redundancy to the structure, and permitting the practice routes to reach the demanded keys along with routing failure tests to perceive the modified malevolent nodes.

There are a number of countermeasures to mitigate the effects of eclipse attacks. The first mitigation method is entitled “Elimination of Far Successors” where every authentic node in the successor list autonomously calculates the distance between every pair of the following nodes and whenever a node is categorized as a malevolent one that equivalent entry is immediately eradicated from the list \cite{rottondi2016detection}. In addition, Bitcoin clients use a classification system to distinguish between new buckets- consisting of a list of all peers and unsupported outgoing contacts with the client; and tried block-containing IP address lists that already made contacts with the client. By employing this system, attackers can be prevented to cause an eclipse attack \cite{aggarwal2021attacks}. Expanding the network size also makes it complicated for an attacker to manage a huge segment of the network and execute the attack. Employing peer-discovery mechanisms is another mitigation technique that helps the nodes to identify and link to a diverse group of peers which facilitates cutting the likelihood of being attacked by a cyber actor.

\begin{figure*}[ht]
\centerline{\includegraphics[width=1.0\textwidth]{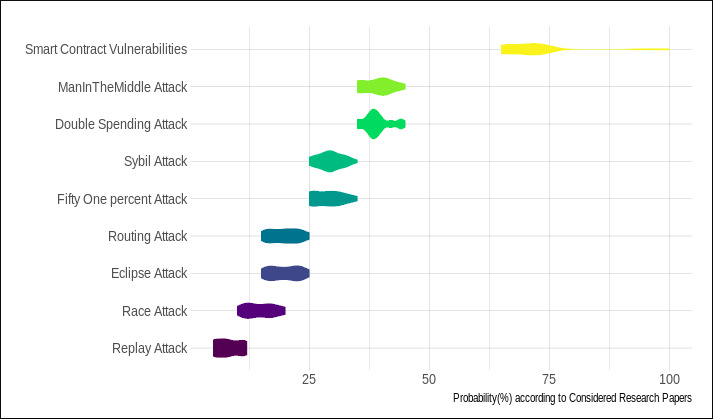}}
\caption{ Probability of blockchain attacks by attack type }
\label{fig:attack-probability}
\end{figure*}

\subsubsection{Double-Spending Attack}
All these thorough research efforts have not only yielded a detection methodology but also established preventative measures to mitigate the risk of double spending. The major detection techniques and defense mechanisms employed for identifying double spending attacks involve verifying the transaction and blocking times while updating the existing block instead of removing it \cite{begum2020blockchain}, continuously monitoring the disconnection period to employ recipient-oriented transaction [\cite{lee2018recipient}, time stamping during the chain building process that augments the number of confirmations establishing a coherent chain  \cite{DBLP:journals/corr/abs-1805-05004}. To detect double spending attacks, \cite{8733108, 8653269} proposed transaction time and block verification adopt the Longest Chain Rule implemented in PoS protocols, Key Evolving Cryptography and Expanding the Number of Confirmed Blocks while \cite{cryptoeprint:2018/236, 10.1007/978-3-662-58820-8_9} emphasize on confirmation of the ledger and monitoring node connections to detect and Disabling the unconfirmed connections and transaction forwarding are the imperative security measures to safeguard the integrity and reliability of the blockchain network \cite{cryptoeprint:2018/236, 10.1007/978-3-662-58820-8_9}.

\subsubsection{Sybil Attack}
In \cite{10.1016/j.future.2017.08.048}, the authors have identified a susceptibility in P2P networks, whereby a malevolent node can be accorded the same treatment as genuine nodes. To address this issue, they have introduced TrustChain, a blockchain solution that leverages a Sybil-resistant algorithm known as NetFlow, which is based on calculating node reputation. Sybil node injection into the Routing Table (also known as the RTI attack) represents yet another vulnerability of blockchain, which can be averted through the implementation of a decentralized registration mechanism \cite{8710078}. Tayebeh et al. highlighted the potential flaws in Bitcoin's Nakamoto consensus protocol, which can be addressed by introducing identity verification during the initial phase to constrain nodes exhibiting abnormally high computational power \cite{DBLP:journals/corr/abs-2002-06531}. M. Quintyne-Collins highlighted the feasibility to establish a connection between transactions and trace the identity of a user or company responsible for each transaction despite the pseudo-anonymous nature of blockchain systems \cite{cryptoeprint:2019/1111}. While it may not be possible to fully eliminate the threat of a linking attack, it can be partially mitigated by implementing measures such as the imposition of non-refundable deposits, the application of time-locks on the utilization of funds, and the use of a coin-age mechanism. The node authentication process of the P2P network during the blockchain network's joining phase poses a significant vulnerability to the system referred to by P. Swathi et. al \cite{8944507}, making it susceptible to the insertion of Sybil nodes that may result in block propagation delays. To address this, it is recommended to introduce a rigorous node authentication process prior to network admission. This could include the implementation of a network joining fee, validation of the source of node connection, and fair forwarding behavior of a node over a predetermined period of time.

\subsubsection{Replay Attack}
Detection techniques for replay attack vulnerabilities in the blockchain involve monitoring for duplicate transactions across both chains resulting from a hard fork or chain split \cite{nfting_what_2022}. By analyzing transaction history in both chains, it is possible to identify any suspicious or repeated transactions that could indicate the presence of a replay attack \cite{duan2022multiple}. Furthermore, implementing security measures like one-time private-public key pairs \cite{dasgupta2019survey}, elliptic curve-based encryption \cite{dasgupta2019survey}, and including variables such as timestamps or nonces in the signed messages \cite{duan2022multiple} can help in detecting and mitigating the risks associated with replay attacks.

Several security measures can be implemented to mitigate replay attacks in the blockchain. One effective solution is implementing strong replay protection, which automatically adds a marker to the new blockchain after a hard fork, ensuring that transactions made on the new blockchain will not be valid on the original chain and vice versa \cite{nfting_what_2022}. Another approach is splitting coins, which involves creating separate and distinct transactions on each chain, thus preventing the possibility of replay attacks \cite{duan2022multiple}. Opt-in replay protection is an additional measure that allows users to manually mark their transactions so that they are no longer valid on the other chain, providing them with more control over the security of their transactions \cite{nfting_what_2022}. Combining these security measures makes it possible to significantly reduce the risk of replay attacks in the blockchain ecosystem.

Table \ref{tab:tab2} depicts a concise overview of blockchain attack detection and mitigation techniques.
\section{Our Findings}
In our research, we came across a number of significant findings that have to do with reducing blockchain attacks and boosting security. These results are essential to our comprehension of the problems and solutions that could be found in this area. We have identified several common forms of attacks on blockchain technology, such as double-spending attacks, 51\% attacks, and weaknesses in smart contracts. These attacks may have serious repercussions, jeopardizing the security and integrity of blockchain networks, leading to monetary losses, and damaging public trust. We have also assessed the various technologies and security measures used to prevent and stop these attacks. Solutions include enhanced consensus algorithms, cryptographic methods, smart contract auditing, and secure development procedures. Furthermore, given that various industries and use cases have different security needs, our research emphasizes the significance of customizing security measures to meet the needs of each blockchain application. As blockchain security is a constantly evolving problem, it is crucial to stay informed about new threats and vulnerabilities. Ongoing research and collaboration are also essential. We can protect our blockchain systems and advance secure blockchain technology by utilizing these findings to develop and implement strong security measures.

According to figure \ref{fig:attack-probability}, we found that the likelihood of various blockchain attacks varies. According to our analysis, there is a 30\% chance of a 51\% attack, a 70\% chance of smart contract vulnerabilities, a 10\% chance of a replay attack, a 20\% chance of a routing attack, a 40\% chance of a double-spending attack, a 30\% chance of a sybil attack, a 40\% chance of a man-in-the-middle attack, a 15\% chance of an eclipse attack, and a 20\% chance of a routing attack. These probabilities give important information about the propensity for each attack to occur. For instance, double-spending attacks and smart contract vulnerabilities were found to have higher probabilities, indicating a higher risk in these areas. We can effectively prioritize our security measures and allocate resources by being aware of these probabilities. We can create targeted defense strategies to lessen the effects of attacks by concentrating on the ones that are most likely to occur. These discoveries advance knowledge of blockchain security as a whole and offer practitioners and researchers helpful pointers for defending blockchain systems against possible threats.

\section{Conclusion}
Our research has offered a thorough analysis of reducing blockchain attacks and boosting security. Because of its decentralized nature and the benefits it provides in terms of transparency, immutability, and security, we have realized the enormous potential of blockchain technology across industries. However, we have also identified some of the major difficulties that blockchain networks face, particularly with regard to security flaws and assaults. The severity and recurrence of these threats have been highlighted by actual attack examples, emphasizing the pressing need for efficient solutions. The significance of addressing blockchain security issues is further highlighted by the rise in reported vulnerabilities and cybercrime. We have determined common forms of attacks on blockchain technology and assessed their effects on security and system integrity by responding to our research questions. Significant threats to blockchain systems have been identified as 51\% attacks, double-spending attacks, and smart contract vulnerabilities. These assaults have the potential to cause sizable financial losses and erode public confidence in technology. Additionally, we have assessed a number of security tools and techniques that can be used to stop and prevent malicious blockchain attacks. We have learned important things about the efficacy of these measures and their limitations through our analysis. We can use this information to inform our decisions about how to implement suitable security measures. Our research also emphasizes how crucial it is to adjust security precautions to each blockchain application's particular requirements. Customized methods are required due to the unique security considerations required by various industries and use cases. In order to ensure the implementation of efficient security measures, we developers and practitioners are aware of the importance of carefully analyzing the unique characteristics and threats of our blockchain systems. Overall, by offering a thorough analysis of attack mitigation techniques, our research adds to the body of knowledge on blockchain security. Our study's conclusions will help us and other researchers make wise decisions and put strong security measures in place. Looking ahead, we understand that ongoing research and cooperation are required to address new threats and vulnerabilities in the dynamic environment of blockchain technology. To further enhance the security and resiliency of our blockchain systems, we will actively investigate cutting-edge methods and technologies. We are dedicated to ensuring the trust, integrity, and broad adoption of blockchain technology by giving security top priority and taking preventative measures. Our research emphasizes the significance of enhancing security and mitigating blockchain attacks. We are confident that by addressing these issues, we can fully realize the potential of blockchain technology and completely transform various industries.

\bibliographystyle{spbasic}
\bibliography{biblography.bib}

\end{document}